# Visualizing Landau levels of Dirac electrons in a one-dimensional potential

Yoshinori Okada[1], Wenwen Zhou[1], Chetan Dhital[1], D. Walkup[1], Ying Ran[1], Z. Wang[1], Stephen D. Wilson[1] & V. Madhavan[1]

Using scanning tunneling spectroscopy, we study a 3D topological insulator $Bi_2Te_3$ with a periodic structural deformation (buckling). The buckled surface allows us to measure the response of Dirac electrons in a magnetic field to the presence of a well-defined potential variation. We find that while the n=0 Landau level exhibits a 12meV energy shift across the buckled structure at 7.5T, the amplitude of this shift changes with the Landau level index. Modeling these effects reveals that the Landau level behavior encodes information on the spatial extent of their wavefunctions. Our findings have important implications for transport and magneto-resistance measurements in Dirac materials with engineered potential landscapes.

[1]Department of Physics, Boston College, Chestnut Hill, Massachusetts 02467, USA

When a magnetic field is applied to a solid, the electrons fall into discrete Landau levels (LLs) represented classically by cyclotron orbits. As energy increases, the increasing spatial spread of the Landau level wavefunctions can be visualized classically in terms of a growing cyclotron orbit radius. Understanding the effects of the spatial size of these wavefunctions is particularly important for inhomogeneous systems where the changing relative size of the cyclotron radius in comparison to the length scale of potential variations can create energy dependent changes in the magneto- transport properties. For example, Weiss oscillations in magnetoresistance measurements of patterned semiconductor two dimensional electron gas systems (2DEGs) have been attributed to a periodic commensurability between the potential periodicity and the cyclotron radius[1, 2, 3]. One of the most powerful methods to probe these effects on a local scale is to measure Landau levels in an inhomogeneous system[4] with scanning tunneling microscopy (STM). Landau levels appear as a sequence of peaks in STM spectra and have been observed in many systems including graphene,[5, 6, 7] conventional 2DEGs[8, 9], and topological insulators[10, 11]. However, although a few experiments have discussed the effects of spatial inhomogeneity on Landau level energies,[7, 9] there have been no experiments directly probing the systematic effects of the increasing wavefunction size on the electronic structure.

We have recently discovered that the prototypical topological insulator $Bi_2Te_3$ is susceptible to a periodic buckling, which results in a periodic potential variation superimposed on the Dirac electrons. The linear dispersion and the well–defined potential variation make this an ideal system to study size effects of Landau level wavefunctions. The ordered nature of the stripes also makes this system ideal for investigating the effects of periodic potentials on the Dirac electrons in topological insulators. The effects of such periodic potentials on the Dirac state have been extensively explored theoretically[12, 13, 14, 15, 16,] and experimentally in graphene[17, 18, 19, 20], but so far there have been no such studies in topological insulators. Importantly, spatial modulations of the Dirac electrons can create fundamentally new boundary modes[21], which are relevant to applications of these materials in quantum computation or spintronics.

$Bi_2Te_3$ samples[22] were studied with an ultra high vacuum (UHV), 4K STM. Single crystal samples were cleaved at room temperature in UHV before being inserted into the STM head. Differential conductance *(dI/dV(r,eV))* maps and spectra were obtained using a standard lock-in technique with bias modulations of 0.4 mV to 2 mV amplitude depending on the data set. STM topography of Bi$_2$Te$_3$ reveals both flat regions (Fig. 1a), as well as regions with 1D modulations (Fig. 1b). Unlike the stripe phases in a 2D electron gas,[23, 24] the width of these modulations (stripes) is large (~500 Å i.e., 1000 Å periodicity) and independent of strength of the magnetic field (Fig. 1c). To study the effect of the periodic potential imposed by the stripes on the Dirac dispersion, we compare the STM spectra in a perpendicular magnetic field obtained in striped and flat regions of Bi$_2$Te. Local LLs have so far been measured in Bi$_2$Se$_3$[10, 11] and Sb$_2$Te$_3$,[25] but they have not yet been reported in Bi$_2$Te$_3$ samples. Figure 2a shows a *dI/dV* spectrum on a flat region at a magnetic field of 7T. Since the LLs are small perturbations on the large background density of states arising from the valence or conduction bands, all LL data is henceforth presented with the background subtracted (Fig. 2b). The energy dependence of the LLs can be calculated in terms of semi-classical theory where it can be shown that in a magnetic field the momentum vector (k-vector) is quantized at the nth LL and is proportional to $\sqrt{nB}$ (n is the LL index and B is the magnetic field). The E-k dispersion then dictates the actual relationship between E and n. The Landau level spectrum can therefore be used to determine the surface state band dispersion in a regime inaccessible to the quasi-particle interference method[5, 11]. When the g-factor is normal (~2), a linear dispersion relation directly translates into a linear dependence of E$_n$ on $\sqrt{nB}$:

$$E_n = E_D + \sqrt{2e\hbar nB v^2_F}$$ where E$_n$ and E$_D$ are the energies of the nth LL and the Dirac point respectively and $v_F$ is the velocity. A plot of E$_n$ against $\sqrt{nB}$ reveals a reasonably good linear fit (Fig. 2d) up to approximately 130 meV above the Dirac point as expected for the dispersion of a topological insulator. In regions away from the impurities, the Landau levels in the flat regions are quite homogeneous (Fig. 2c). Interestingly, we find a new peak (labeled n') between the n=0 and n=1 Landau levels which also appears in the magnetic field. An analysis of the possible origins of this peak is not relevant to the main focus here, and we do not discuss this feature any further in this paper.

We now study the effect of the periodic potential on the Landau level behavior of the Dirac electrons. As discussed theoretically in graphene, the well-defined linear dispersion of the surface state makes this system ideal for separating magnetic field effects from band structure effects. We find that the observed topographic stripes have a dramatic effect on the LL behavior. Figure 3 shows a plot of the LL energies as we traverse the stripes. Our first observation is that the n=0 and 1 LL energies follow the topography (purple line in Fig. 3b) remarkably closely. This is clear evidence that the topographic stripes impose a smoothly varying 1D potential on the Dirac electrons. In fact, using these levels as a probe, we can attribute an energy scale of approximately 12 meV to the peak-to-peak potential difference. As we progress to higher index levels however, we find that their energies in the crests and troughs are no longer offset by a constant amount. This behavior is clearly reflected in figure 3d and e where we compare the LLs at the extreme positions in valley and crest of the stripes.

The simplest interpretation of Landau level spectra is that LL energies follow the surface state dispersion[5, 11], which implies that the stripes induce energy dependent changes in the dispersion. This is however contradictory to earlier studies of the effects of potential variations arising from doping inhomogeniety, which create simple energy independent, rigid shifts of the bulk and surface bands[26]. Given this inconsistency, we consider the possibility that our observations are a consequence of the increasing spatial spread of the Landau level wavefunctions. As we show in the analysis and discussion below, this will be a particularly important consideration in inhomogeneous samples with potential variations due to ripples, dopants or other defects.

The consequences of the increasing spread of the Landau level wavefunctions with LL index can be simplistically visualized as an averaging effect over sequentially larger areas of real space. The concept that the Landau levels cannot sample potential variations on length scales smaller than the cyclotron radius has been invoked to explain LL behavior in 2D electron gas systems,[9] but the background potential was random and unknown in those cases. Additionally, there were added complications due to band bending. In our case, with the potential variation quite well defined by the periodic stripes and a surface state dispersion that is well known, we are in a powerful position to visualize the effects of the wavefunction size on the Landau level spectra. In order to analyze our data we chose the Landau gauge which offers the most suitable description according to the symmetry of this system. For a parabolic band in the Landau gauge, the Hamiltonian looks like that of a harmonic oscillator with the Landau level index (n) taking on integer values and labeling the successively higher energy states in the harmonic oscillator potential. For a Dirac surface state however, the eigenstates are exactly $\psi_{|n|}(x,k_y) \otimes |\uparrow\rangle$ for n=0 and linear superpositions of $\psi_{|n|}(x,k_y) \otimes |\uparrow\rangle$ and $\psi_{|n-1|}(x,k_y) \otimes |\downarrow\rangle$ when $n \neq 0$ (see supplementary information[27] for details). For the purposes of an initial estimate of the size of the nth LL, we calculate the size of the larger of the two i.e. the spatial extent of $\psi_{|n|}(x,k_y) \otimes |\uparrow\rangle$ in the x-direction, which is well represented by: $\sqrt{\langle x^2 \rangle} = \sqrt{(2n+1)\hbar/eB}$. For a 7T field the size of the n=1 LL wavefunction (~27 nm) is well within a single (bright or dark) region of the stripes. At n=3 however, the size increases to ~60 nm. This simple analysis indicates that in our case where the stripe width is ~50nm, the effects of the increasing wavefunction size may be substantial above n=3.

In order to shed further light on the size effects of the LL wavefunctions, we perform a simulation of Landau level spectra. Here we assume that to first order, the surface state dispersion follows the smoothly varying potential due to the stripes (Fig 3 and supplementary information[27]) and use the unperturbed Dirac Landau level wavefunctions discussed above to calculate the resultant STM spectra. The simulation also takes into account spatially adjacent Landau level states. We find that this is important to accurately capture many details of the LL variations across the stripes. Our analysis provides a simple method to take the LL wavefunctions into account in analyzing transport or quantum hall measurements of samples with engineered potential landscapes. Importantly, depending on the size of the wavefunction relative to the length scale of the potential variation (Figs. 4b and 4c), the electronic and transport properties of the system are expected to show distinct regimes of behavior[28, 29].

Certain aspects of the relative intensity of the LL peaks are however not completely captured by the simulation. These provide clues to possible interaction effects in this system. First, the intensity of the n=0 peak is particularly weak in our data. While the reasons for this are not well understood, one possibly is that the n=0 LL is affected by its proximity to the bulk valance band at the $\Gamma$ point in $Bi_2Te_3$[30]. Also, the intensity differences between B and D spectra for higher index LLs as well as the oscillatory intensity in the B region as a function of n for the higher Landau levels are not completely captured by our model. While there is some oscillation in Landau level intensity in our simulations at higher indices as can be observed in figure S4b, it does not reproduce the trends seen in the data. However our model calculation does not take electron-electron or electron-phonon interactions into account, which are expected to affect the Landau level intensity. For example, the sharp rise in intensity close to the Fermi energy has been attributed to a decrease in electron-phonon coupling at these energies[10, 11]. More sophisticated calculations are necessary to capture the interaction effects between the surface state and bulk bands as well as electron-phonon effects which are most certainly energy dependent.

Interestingly, the behavior of the lower index Landau level as shown in Fig 3c implies the existence of alternating spatial regions (along the crests and troughs of the stripes respectively) where the Landau level energies oscillate. If the Fermi energy were moved to the middle of one of these oscillating Landau levels (the n=1 LL for example), the crest and troughs would represent regions with alternating filling fractions, as depicted schematically in Fig. S7 of the supplemental information[27]. Similar to the edge states in quantum Hall effect, the boundaries between the crests and troughs would then be host to gapless chiral metallic modes[31] (see supplementary information[27] for a simple mathematical picture of the chiral modes). These chiral modes bound to the domain walls of the adjacent stripes are ideal one-dimensional quantum wires, which are stable to disorder due to the topological nature of the surface state band. To complete the conditions for realizing the 1D quantum wires however, the Fermi energy must first be moved closer to the lower index levels, which can potentially be achieved by gating or doping these systems. Since the stripes maintain their unidirectional nature over micrometer distances, this renders the resulting quantum wire states potentially useful for electrical and thermal transport applications.

In summary, we have used STM spectroscopy to measure the Landau levels in a topological insulator with a one-dimensional structural buckling of approximately 100nm periodicity (stripes), which imposes a ~12 meV periodic potential on the Dirac electrons. Although the linear dispersion of the Dirac electrons dictates a linear dependence of the LL energies on $\sqrt{nB}$, the LLs in the striped regions show marked deviations from this picture. By simulating the LLs in a periodic potential and taking account the quantum mechanical wavefunctions for the different indices, we were able to capture the unexpected LL behavior. Our model provides a simple way to take the spatial extent of the LL wavefunctions into consideration for describing LL behavior in spatially inhomogeneous systems. Finally, our experiments open up the door to theoretical and experimental investigations of the response of Dirac electrons in topological insulators to periodic potentials.

**Acknowledgements:**

The authors sincerely acknowledge helpful discussions with Dr. Markus Morgenstern and Hsin Lin. The authors acknowledge support by US NSF-CAREER-0645299 and DOE DE-SC0002554.

Fig. 1 Topography of flat and stripes regions

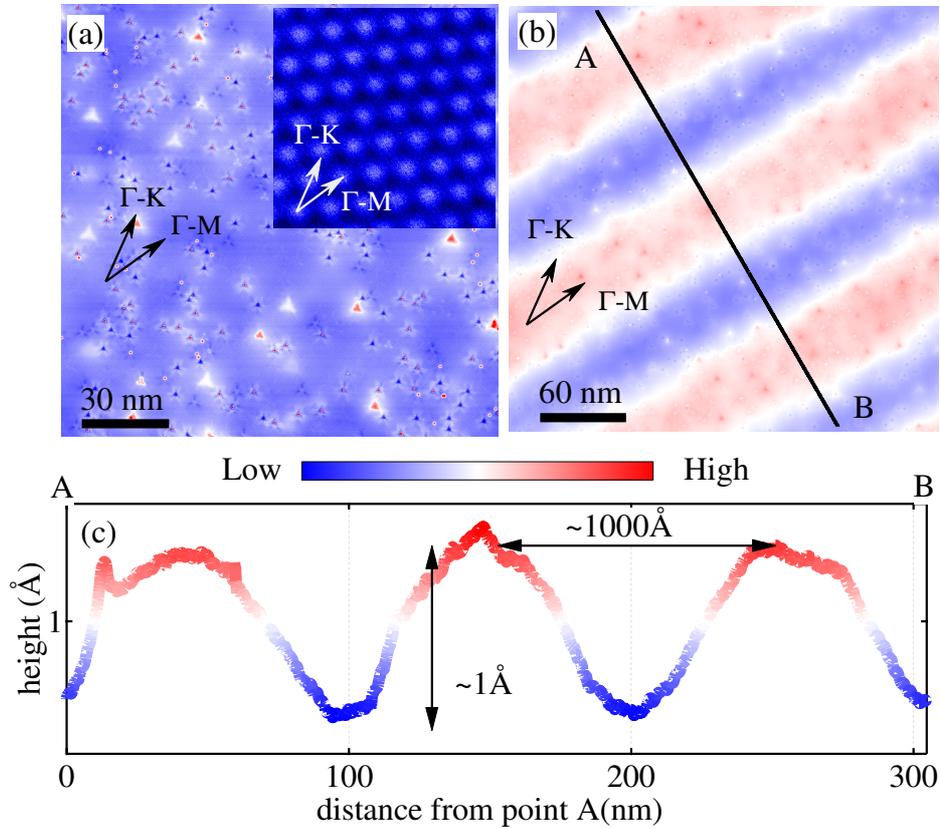

**Fig. 1** Topography of flat and striped regions **(a)** Topography of a flat region **(b)** 300 nm topography of striped region in the same sample obtained with a +200mV bias voltage. The stripes are aligned to within 5 degrees of one of the high symmetry directions of the crystal and remain contiguous over micron length scales. Both striped and flat (no topographic stripes) regions exist within each sample. The stripes occupy 5-10 µm-sized patches separated by the flat areas. The small sharp spikes in the striped topography are the naturally occurring impurities in the sample. **(c)** Linecut across stripes showing one-dimensional topographic modulations with ~1000 Å periodicity and ~ 1Å height. Stripes were observed on multiple samples with multiple tips.

Fig. 2 Landau Levels in flat regions

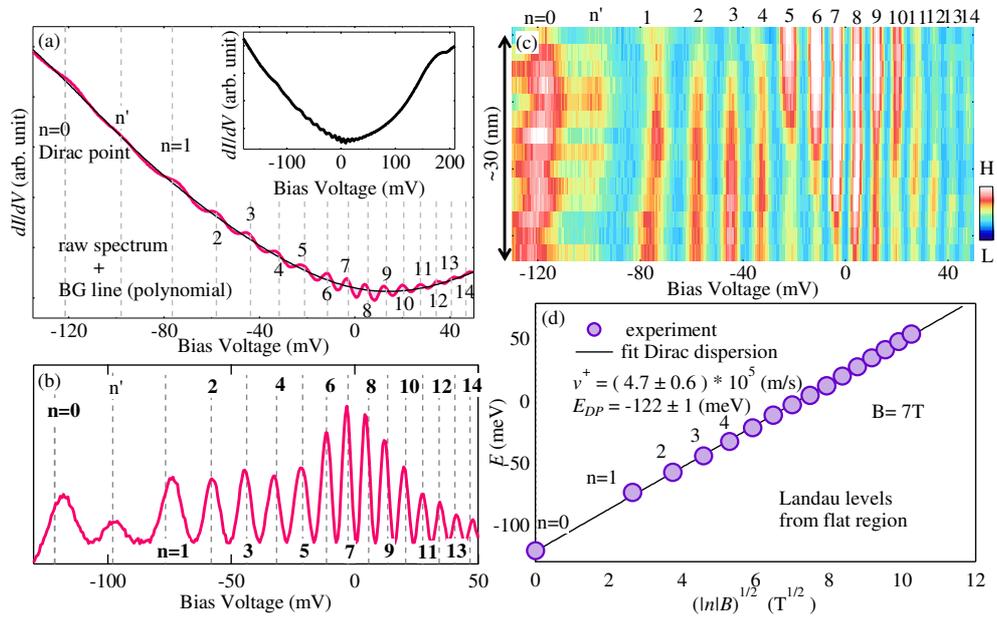

**Fig. 2** Landau levels in flat regions **(a)** *dI/dV* spectrum (see inset for full spectrum) at 7T with a polynomial fit to the background **(b)** dI/dV spectrum with the background subtracted. **(c)** Intensity plot of background subtracted *dI/dV* spectra across a flat surface away from impurities. **(d)** Peak positions of the Landau levels shown in **b.**

Fig. 3 Variation of Landau level energy across periodic potential

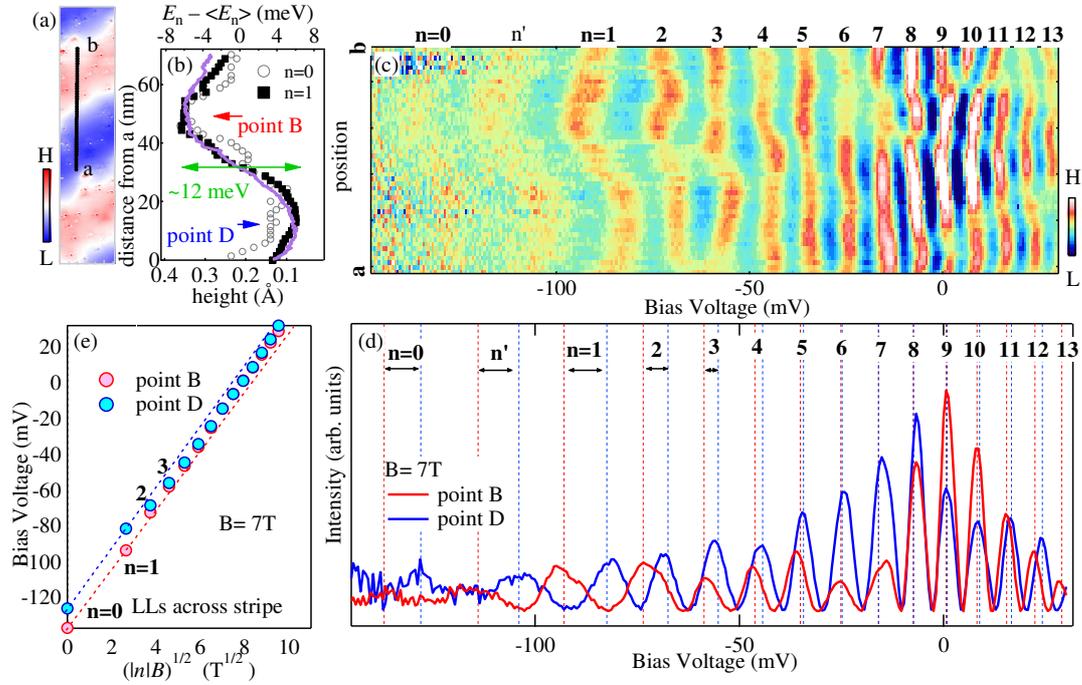

**Fig. 3** Variation of Landau level energy across stripes. **(a)** and **(b)** Topography and corresponding topographic height variation (purple) across the stripes along the line from point a to point b. Also shown is the Landau level energy variation across the same linecut for the n=0 (gray circles) and n=1 (black squares) levels. **(c)** Intensity plot of background subtracted *dI/dV* spectra as we traverse the stripe from point a to point b. **(d)** Comparison of LLs at the crest (point B) and the valley (point D) of the stripes, showing the relative shift in the LL energies. The dotted lines refer to the peak positions plotted in e. **(e)** Peak positions of the LLs shown in **d.**

Fig. 4: Simulation

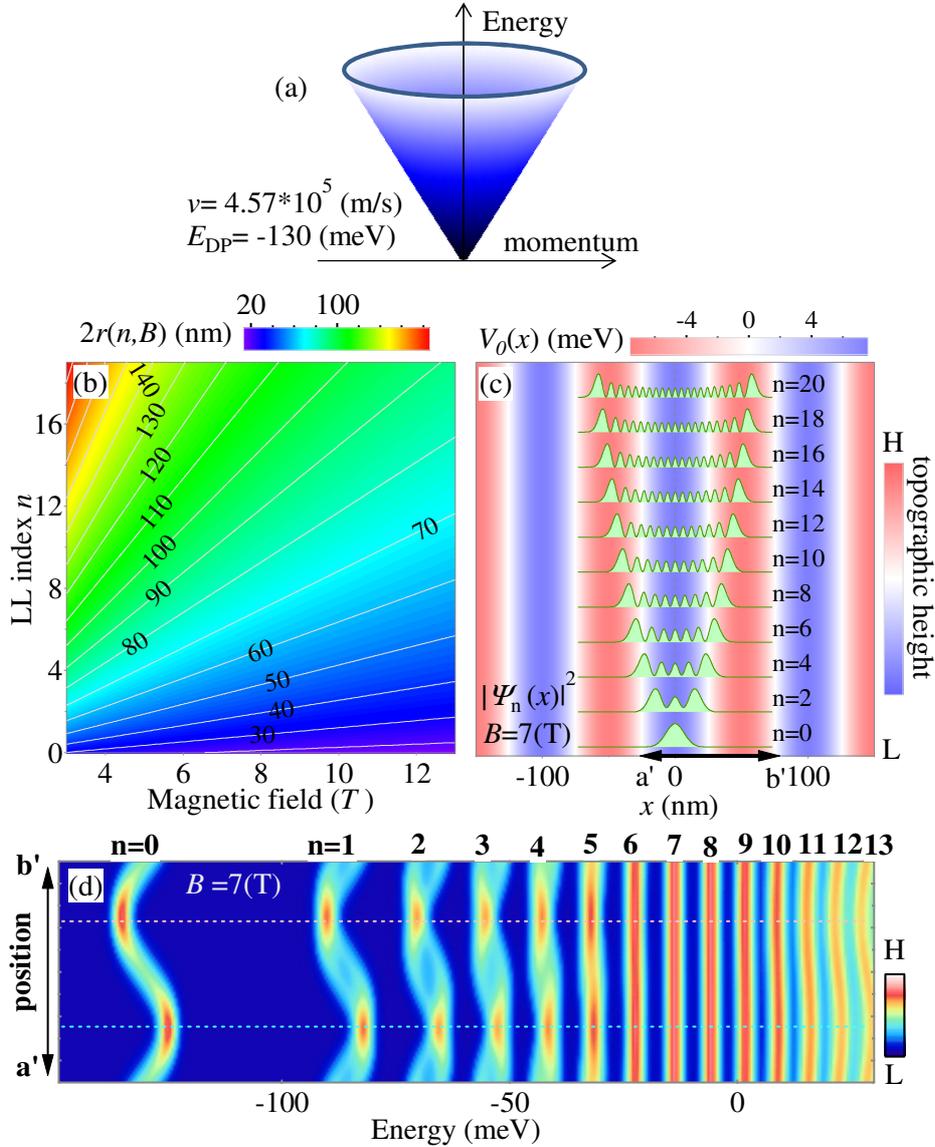

**Fig. 4** Simulation of Landau level energy across stripes. **(a)** Simulation parameters and model for results shown in d. $V_0(x)$ is the sinusoidal potential variation superimposed on the Dirac electrons. **(b)** Intensity plot of change in size of LL wavefunction represented by r (n,B)=$\left\langle\sqrt{x^2}\right\rangle$ with magnetic field and LL index **(c)** Wavefunction squared for Dirac electrons in a magnetic field at 7T superimposed on periodic stripes of 100nm periodicity showing the relative length scales. **(d)** Simulated intensity plot of *dI/dV* spectra across the stripes using the model described in the supplementary information[27].